\documentclass[floatfix,aps,prl,twocolumn,superscriptaddress]{revtex4-2}

\usepackage{graphicx}
\usepackage{bm}
\usepackage{color,soul}
\usepackage{amsmath,amsfonts}
\usepackage{comment}
\usepackage{ulem}
\usepackage[english]{babel}

\usepackage{etoolbox}
\apptocmd{\sloppy}{\hbadness 10000\relax}{}{}

\renewcommand{\emph}{\textit}
\newcommand{\beq}{\begin{equation}}
\newcommand{\eeq}{\end{equation}}

\begin{document}

\title{Anisotropic Melting of Frustrated Ising Antiferromagnets} 

\author{Matthew W. Butcher}
\affiliation{Department of Physics \& Astronomy, Rice University, Houston TX 77005, USA}

\author{Makariy A. Tanatar}
\affiliation{Ames Laboratory, Ames, IA 50011}
\affiliation{Department of Physics \& Astronomy, Iowa State University, Ames, IA 50011}

\author{Andriy H. Nevidomskyy}
\email{nevidomskyy@rice.edu}
\affiliation{Department of Physics \& Astronomy, Rice University, Houston TX 77005, USA}

\date{\today}

\begin{abstract}
    Magnetic frustrations and dimensionality play an important role in determining the nature of the magnetic long-range order and how it melts at temperatures above the ordering transition $T_N$.
    In this work, we use large-scale Monte Carlo simulations to study these phenomena in a class of frustrated Ising spin models in two spatial dimensions.
    We find that the melting of the magnetic long-range order into an isotropic gas-like paramagnet proceeds via an intermediate stage where the classical spins remain anisotropically correlated. 
    This correlated paramagnet exists in a temperature range $T_N < T < T^\ast$, whose width increases as magnetic frustrations grow.
    This intermediate phase is typically characterized by short-range correlations, however the two-dimensional nature of the model allows for an additional exotic feature -- formation of an incommensurate liquid-like phase with  algebraically decaying spin correlations. 
    The two-stage melting of magnetic order is generic and pertinent to many frustrated quasi-2D magnets with large (essentially classical) spins.

\end{abstract}

\maketitle 

The formation of long-range magnetic order (LRO) upon cooling from a disordered paramagnetic (PM) phase is akin to a gas-to-solid transition. The ``solid'' phase is characterized by a spontaneously broken symmetry with long-range order in the spin-spin correlations. 
Studies of geometric frustrations  and their effect on this transition have a long history -- it is well established that frustrations suppress the N\'eel transition temperature $T_N$ relative the Curie-Weiss temperature, which is often quantified by the Ramirez frustration ratio $\eta=T_{CW}/T_N$~\cite{ramirez_strongly_1994}. In some highly frustrated lattices, such as the corner-sharing tetrahedra in pyroclore magnets~\cite{gardner_magnetic_2010}  or corner-sharing triangles in kagom\'e compounds~\cite{huse_classical_1992, mendels-kagome-review}, $T_N$ is suppressed to zero, with the formation of a (classical~\cite{chalker-frustrations-review} or quantum~\cite{QSL-review}) spin liquid and associated order-by-disorder lifting of the extensive ground state degeneracy~\cite{moessner-chalker}. 
However, highly frustrated geometries are not the only way to suppress LRO. Instead, one may consider seemingly simple bipartite lattices, such as the square or cubic lattice systems with competing spin interactions between the nearest and farther neighbors (a paradigmatic $J_1-J_2$ model on a square lattice is one such example~\cite{misguich_review_2005}).
Apart from suppressing the N\'eel temperature, do such interaction-induced frustrations affect the process of ``melting'' of the LRO when the ground state is not extensively degenerate?
Furthermore, what is the role of the dimensionality of the magnetic system -- which can be controlled in principle by tuning the degree of anisotropy of the exchange interactions along the different crystal directions -- on the strength of thermal fluctuations? 

In this Letter, we perform Monte Carlo simulations to study a \textit{classical} magnet with anisotropic interactions to elucidate the effect of both the anisotropy and magnetic frustrations on the process by which the magnetically ordered ``solid" melts. Our key finding is that this melting proceeds via an intermediate stage in a range of temperatures above the N\'eel ordering temperature $T_N$, where the correlations between the spins remain significant and retain the `knowledge' of the anisotropy present in the Hamiltonian. This intermediate correlated paramagnet (CPM) eventually undergoes a crossover into a more conventional paramagnet at a temperature $T^\ast > T_N$, above which the correlation length is of the order lattice constant in all directions, and the anisotropy is lost. We find that the dynamic temperature range of this CPM ($T^\ast - T_N$) grows with increasing frustrations, and that the CPM occupies a significant portion of the phase diagram even as $T_N$ is suppressed to zero by frustrations. These classical results are relevant to many experimental systems (both itinerant and insulating magnets) with large magnetic moments that can be treated as classical, where the evidence of the crossover scale $T^\ast$ and the anisotropic CPM has accumulated, for instance in Eu-based helimagnets~\cite{EuCo2A2,EuCo2P2,EuMg2Bi2}, some heavy fermion compounds with helical order~\cite{CeRhIn5entropy,PaglioneRh,CePt}, and in layered ferromagnets~\cite{dim-crossover_CrSiTe3_2019}.

Our second finding is that in certain classical models, such as the anisotropic next-nearest neighbor Ising (ANNNI) model, a correlated paramagnet can exist not just as a crossover region above $T_N$, but as a well-defined phase of its own, characterized by the algebraically (rather than exponentially) decaying spin correlations. As such, this phase is more akin to a liquid with quasi-long-range correlations rather than a gas.
In the case of the two-dimensional (2D) ANNNI model, the existence of such a `liquid' phase was proposed in the 1980s~\cite{bak_commensurate_1982}, and here we demonstrate, through large-scale classical Monte Carlo simulations, its appearence in anisotropic ANNNI model and trace its connection to the aforementioned CPM.

\begin{figure}
\centering
    \includegraphics{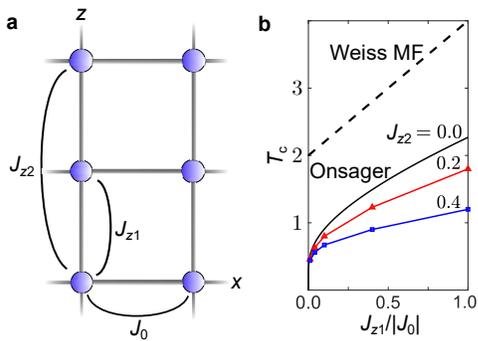}%
    \caption{\label{fig:model_and_mean_field}
    Suppression of $T_N$ with the addition of frustrating interactions.  
    \textbf{(a)} Diagram of interactions in the 2D ANNNI model. 
    \textbf{(b)}   Comparison of $T_N$ from Weiss molecular field theory (dashed line), the Onsager solution ($J_{z2} = 0$, solid black line), and the 2D ANNNI model with $J_{z2} = 0.2 J_{z1}$ (red line) and $J_{z2} = 0.4 J_{z1}$ (blue line).}
\end{figure}

\noindent
\textit{Model} -- The three-dimensional (3D) ANNNI model, with ferromagnetic interactions in the plane and competing interactions along the $c$-axis of hexagonal crystal,  was first proposed by Elliot~\cite{elliott_phenomenological_1961}  to explain complex helicoidal orders and their temperature evolution in the rare-earth magnets. As we show in this Letter, this uniaxial frustration plays a
pivotal role in how this anisotropic magnet `melts'. 
We consider as our starting point the 2D ANNNI model  (Fig.~\ref{fig:model_and_mean_field}a): 
\begin{equation}\label{eq:ANNNI2D}
    H = \sum_i \left[J_0 \sigma_i \sigma_{i+\hat{x}} + J_{z1} \sigma_i \sigma_{i+\hat{z}} + J_{z2} \sigma_i \sigma_{i+2\hat{z}}\right].
\end{equation}
The variables $\sigma_i\in\{-1,1\}$ are classical Ising spins, with ferromagnetic coupling along $x$ direction $J_0 < 0$, and competing antiferromagnetic interactions along $z$: \mbox{$0 < J_{z2} < J_{z1}$.} 
The reason for our considering a 2D version of the model is both because of its simplicity, and because it evades the entropy-induced multitude of incommensurate phases (``devil's staircase")
experimentally found in CeSb~\cite{CeSb} and specific to the 3D ANNNI model at finite temperatures~\cite{von_boehm_devils_1979,bak_ising_1980,fisher_infinitely_1980,selke_two-dimensional_1980,bak_chaotic_1981,bak_commensurate_1982,pleimling_anisotropic_2001,murtazaev_monte-carlo_2009}.

First, we consider anisotropy in the absence of frustration, in this case with $J_{z2} = 0$.  The Weiss molecular field approach~\cite{weiss_hypothese_1907} predicts an ordered phase of ferromagnetic chains that are stacked antiferromagnetically in the $z$-direction, occuring below a mean-field critical temperature given by $k_B T_{MF} = 2(|J_0| + |J_{z1}|)$ (Fig.~\ref{fig:model_and_mean_field}b, dashed line).  The thermal fluctuations alter this behavior considerably, as famously shown by Onsager~\cite{onsager_crystal_1944}, resulting notably in the much lower transition temperature $T_N$ than predicted by Weiss molecular-field theory, namely given by the following equation (with $\beta_c = (k_B T_N)^{-1}$): 
\begin{equation}
    \sinh(2\beta_c |J_0|)\sinh(2\beta_c|J_{z1}|) = 1.
\end{equation}
The resulting critical temperature is shown in Fig.~\ref{fig:model_and_mean_field}b (solid black line) as a function of the anisotropy $J_{z1}/J_0$. Notably, as a result of thermal fluctuations, $T_N \rightarrow 0$ in the one-dimensional (1D) limit of $J_{z1}\rightarrow 0$. This effect bears a striking resemblance to the suppression of $T_N$ in the quantum Heisenberg model in two dimensions, which is itself a consequence of the Mermin-Wagner theorem~\cite{mermin_absence_1966}.  The existence and proximity of a lower critical dimension in both the quantum and classical cases is important for emergence of the anisotropic CPM phase we discuss in this Letter.

\begin{figure*}[t!]
\centering
    \includegraphics[width=2\columnwidth]{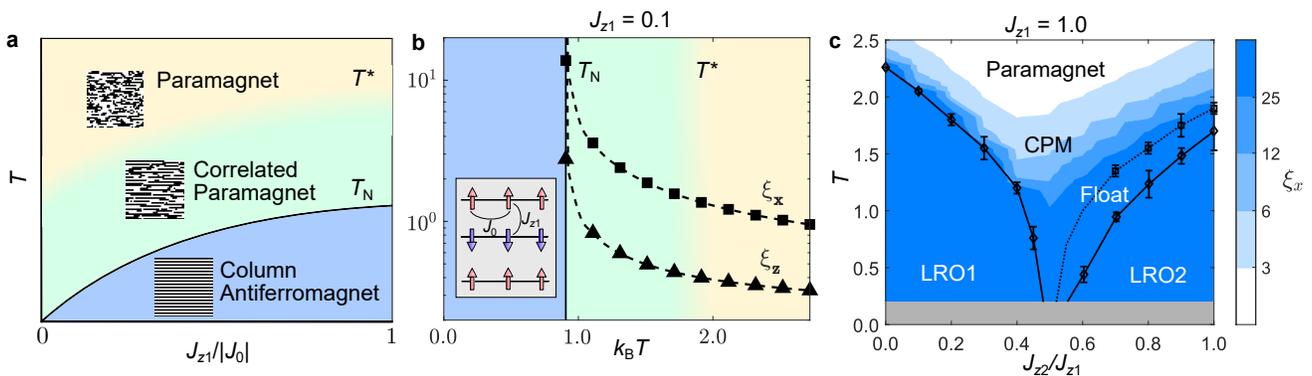}%
    \caption{\label{fig:dimensional_crossover} 
    Anisotropic melting in 2D Ising models.
    \textbf{(a)} Schematic diagram demonstrating the concept of anisotropic melting in A-type antiferromagnet without frustration ($J_{z2} = 0$), a 2D Ising system with spatially anisotropic interactions , $0 < J_{z1} < |J_0|$.  There is an extended region in the phase space at temperatures between $T_N$ and the dimensional crossover scale $T^*$ with qualitatively distinct anisotropic magnetic liquid phase.  
    Insets in (\textbf{a}) are bitmap images of microstates from a Monte Carlo simulation where black squares represent $\sigma_i = +1$ and white represent $\sigma = -1$.  The anisotropic magnetic liquid is visually discernible from the paramagnet.
    \textbf{(b)} The correlation length in the two directions $x$ and $z$, fit from the Onsager solution \cite{onsager_crystal_1944,montroll_correlations_1963} for $J_{z1} = 0.1|J_0|$, with $\xi_x$ becoming appreciable below $T^*$ while $\xi_z$ remains negligible. The inset in (b) illustrates the ordered state with anisotropic interactions.
    \textbf{(c)} Contour plot of the x-direction correlation lengths through parameter space in the 2D ANNNI model for $J_{z1} = |J_0|$.  The $T^*$ scale, defined here as the temperature below which $\xi_x$ grows larger than $3$ lattice sites, is most suppressed when approaching the frustration point $J_{z2}/J_{z1} = 0.5$.
    }
\end{figure*}

\noindent
\textit{Geometric Frustrations and anisotropic CPM} -- With the addition of geometric frustration in the form of an antiferromagnetic coupling between second-neighbour layers $0 < J_{z2} < J_{z1}/2$, the mean-field energy scale $k_B T_{MF} = 2(|J_0| + J_{z1} - J_{z2})$ is lowered slightly, whereas the true transition temperature is further suppressed by frustrations as shown in Fig.~\ref{fig:model_and_mean_field}b (red and blue lines).

Results for the frustrated 2D ANNNI model were computed using classical Monte Carlo simulation with conventional Metropolis updates, as well as cluster updates, and parallel tempering~\cite{Supplement}. By analyzing the spin-spin correlation functions in our Monte Carlo simulations
\begin{equation}\label{eq.correlations1}
    C(\mathbf{r}) \equiv \langle \sigma(0)\sigma(\mathbf{r})\rangle \sim f(\mathbf{r})\cos(q z).
\end{equation}
we extract the correlation lengths $\xi_x$ and $\xi_z$ from the spatial decay of  $f(r)\sim r^{-1/2}\exp(-\sqrt{(x/c)^2 + z^2}/\xi_z)$, where $c=\xi_x/\xi_z>1$ due to anisotropy. Here the factor $r^{-1/2}$ arises in accordance with the canonical Ornstein-Zernike form for correlations in two dimensions~\cite{campanino_ornstein-zernike_2003}. As expected, we find that the correlation lengths
diverge below $T_N$ but are finite and anistropic above the transition as shown in Fig.~\ref{fig:dimensional_crossover}b. This resembles the situation in a classical liquid, which has a short-range order (SRO) but no long-range order. However unlike in a classical liquid, spin correlations retain the ``memory" of the anisotropy $J_{z1}/J_0 < 1$, with the resulting correlation length being much shorter in the $z$-direction $\xi_z \ll \xi_x$ as shown in Fig.~\ref{fig:dimensional_crossover}b. 
One can picture this anisotropic correlated paramagnet (CPM) as consisting of oblong droplets of size $\xi_x \times \xi_z$, with the spins correlated within the droplet but not between them (see the insets in Fig.~\ref{fig:dimensional_crossover}a for Monte Carlo snapshots). As mentioned in the introduction, indications of such an anisotropic CPM are seen, for instance in magnetic scattering measurements above $T_N$ in EuCo$_2$P$_2$~\cite{TanatarEuCo2P2} and at a field-tuned quantum critical point in CeCoIn$_5$~\cite{PaglioneQCP,nonvanishing,Science}.
As the temperature increases, these oblong droplets shrink until eventually their large axis ($\xi_x$) becomes comparable to the lattice spacing -- at that temperature, which we denote by $T^*$, a crossover into a conventional paramagnet occurs, with very short-ranged correlations in both directions. 
Crucially, we find that while magnetic frustrations suppress $T_N$, which is well known, the dynamic temperature range $T_N < T < T^*$  where the anisotropic CPM exists grows with increased frustrations, as shown by the color scale in Fig.~\ref{fig:dimensional_crossover}c. This range becomes especially pronounced when frustration is largest near $J_{z2}/J_{z1}=1/2$.

\begin{figure}
\centering
    \includegraphics[width=\columnwidth]{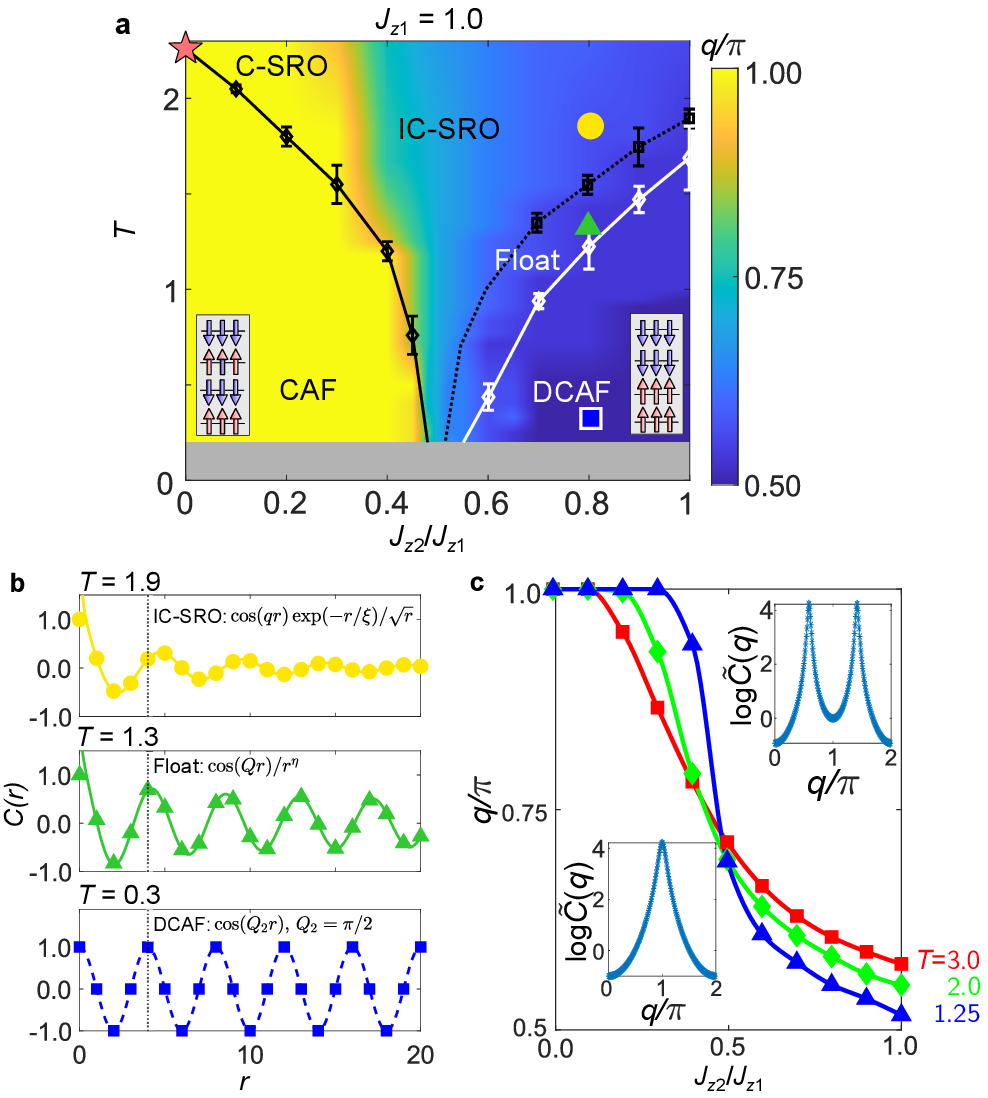}%
    \caption{Incommensurate phases and correlation functions.
    \textbf{(a)} False color plot indicating the evolution of the magnetic ordering wavevector as determined from Monte Carlo simulations with $J_{z1} = |J_0|$.  The Fourier-space correlations $\tilde{C}(\mathbf{q}) \equiv \sum_i e^{-i\mathbf{q}\cdot\mathbf{r}_i} \langle \sigma(0)\sigma(\mathbf{r}_i)\rangle$ are peaked at a wavevector $\mathbf{q} = (0, q)$.  Insets in panel (a) indicate the $q=Q_1$ (CAF, left) and $q=Q_2$ (DCAF, right) low-temperature ordered phases.  
    The star on the vertical axis indicates the critical temperature from the analytical results for the nearest-neighbor Ising model~\cite{kramers_statistics_1941,onsager_crystal_1944}. 
    \textbf{(b)} Monte Carlo data and analytical fits to the spin correlation function for three different temperatures.
    The magnetic wavevector $q$ changes continuously with decreasing temperature across the three panels, becoming $q=Q_2=\pi/2$ in the long-range ordered DCAF phase (bottom graph). The vertical dotted line is a guide to the eye indicating the magnetic period in the DCAF phase. 
    \textbf{(c)} Evolution of $q$ at fixed temperatures as a function of $J_{z2}/J_{z1}$.  The Lifshitz transition gets sharper with lowering of the temperature.
    The insets (b) depict the single- and double-peak structure of the Fourier-space correlations in the CAF and DCAF phases, respectively.}
    \label{fig:phase_color}
\end{figure}

\noindent
\textit{Floating `liquid' phase} -- The aforementioned evolution of a CPM into a paramagnet at $T^\ast$ is  a crossover rather than a true phase transition in the regime $J_{z2}/J_{z1} < 1/2$ where the spin correlations remain commensurate with the lattice ($q=Q_1=\pi$ in Eq.~\eqref{eq.correlations1}). Upon further increase of $J_{z2}$, the correlations in the high-temperature paramagnet  become incommensurate, with the value of $\pi/2 < q <\pi$ shown by color in Fig.~\ref{fig:phase_color}(a). Surprisingly, in this regime the correlated paramagnet acquires a very different character, with spin correlations  that decay algebraically while oscillating with an incommensurate wavevector $q = \pi/2 + \Delta q$:
\beq
C_\text{float}(r) \sim \frac{1}{r^\eta}\cos(q z),
\eeq
shown by the green triangles in Fig.~\ref{fig:phase_color}(b).
Such algebraic correlations are typically seen only at a critical point, whereas here they are a signature of a phase of matter in the extended temperature range $T_N < T < T_\text{fl}$, with the exponent $0<\eta<1/4$ varying smoothly as a function of temperature.
The appearance of such a phase in the 2D ANNNI model was first suggested by Bak~\cite{bak_commensurate_1982}, and later confirmed by several studies~\cite{finel_two-dimensional_1986,shirahata_infinitesimal_2001,chandra_floating_2007-classical}. It is historically called a ``floating'' phase, to do with the appearance of a similar phenomenon in the physics of an incommensurate adsorbent on top of a crystalline substrate~\cite{Coppersmith_1981, Coppersmith_1982, Fisher_1982}.

Our analysis shows that the floating phase is separated from the ordered commensurate phase with $\mathbf{Q}_2 = (0, \frac{\pi}{2})$ (sometimes called the double-column antiferromagnet, DCAF) by a true phase transition at $T_N$, at which the degree of incommensurability of the floating phase vanishes as a power-law: $q-Q_2 = \Delta q \sim (T-T_N)^\beta$ (see Fig.~\ref{fig:phase_color}c). This transition, first investigated by Pokrovskii \& Talapov~\cite{pokrovskii_theory_1980}, is expected to have $\beta=1/2$ in 2D (See SM~\cite{Supplement}).
The physical picture is that above $T_N$, the domain walls proliferate along the $z$-direction, destroying the true long-range order and resulting in the incommensurability of the floating phase.
At the upper boundary $T_\text{fl}$ (dashed line in Fig.~\ref{fig:phase_color}a), the floating phase is separated from the disordered paramagnet (with exponentially decaying correlations, shown by yellow cirles in Fig.~\ref{fig:phase_color}b) by a Berezinskii--Kosterlitz--Thouless (BKT) transition. While Fig.~\ref{fig:phase_color}(a) shows the results for the isotropic case ($J_{z1}/J_0=1$), the same phenomena are observed at arbitrary value of $J_{z1}$, with the main effect of the anisotropy to lower the temperature scales $T_N$ and $T_\text{fl}$ (see SM~\cite{Supplement} for data at $J_{z1}/J_0=0.1$).

\begin{figure}
\centering
\includegraphics[width=0.85\columnwidth]{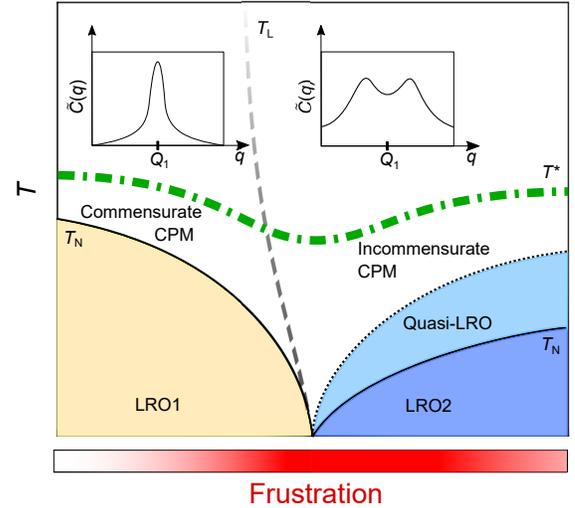}%
\caption{\label{fig:phase_diagram} 
Schematic phase diagram of a frustrated magnet with two competing orders.
The two ordered phases (LRO1, LRO2) are realized below the N\'eel temperature $T_N$, which is suppressed when frustrations are  largest in the center of the horizontal axis. 
At high temperatures, the Lifshitz transition $T_L$ is a crossover from a commensurate $\mathbf{q} = (0, \pi)$ PM to an incommensurate PM with $\mathbf{q} = (0, q)$ as frustration increases.  The insets depict the spin correlation function $\tilde{C}(q)$, roughly proportional to the static spin structure factor measured in neutron scattering. This function
has a single peak in the commensurate phase and double peaks in the incommensurate phase~\cite{schollwock_onset_1996}.  As temperature is lowered, there is a crossover scale $T^*$ below which the in-plane ferromagnetic correlations grow and the system becomes effectively one-dimensional (or two-dimensional, in 3D models). In 2D, the right hand side often contains a quasi-LRO floating phase with power-law spin correlations, which is separated by a BKT phase transition (dotted line) from the incommensurate paramagnet. The true second order phase transition into an ordered magnet occurs at the lower temperature $T_N$ (solid line).  Note that the strength of frustrations, indicated by the intensity of the red bar, does not vanish at the right boundary.}
\end{figure}

\noindent
\textit{Lifshitz transition} --
As the above analysis and Fig.~\ref{fig:phase_color}(a) illustrates, the wavevector $\mathbf{q}$ characterizing the spatial dependence of the spin correlation function can change as a function of temperature. More generally, $\mathbf{q}=(0,q)$ also changes as a function of the ratio $J_{z2}/J_{z1}$, as shown in Fig.~\ref{fig:phase_color}(c), from the commensurate $\mathbf{q}=\mathbf{Q}_1=(0,\pi)$ in the low-frustrated region (yellow in  Fig.~\ref{fig:phase_color}a), to incommensurate at higher frustration. 
This is known as the \textit{Lifshitz transition} (a misnomer, as it is a crossover in the thermodynamic sense), characterized by the appearance of the double-peak structure in the spin-structure factor $S(\mathbf{q})$, as shown in the inset of Fig.~\ref{fig:phase_diagram}. 
Upon crossing the Lifshitz transition, shown schematically by a dashed grey line in Fig.~\ref{fig:phase_diagram}, the wavevector changes continuously away from $\mathbf{q} = \mathbf{Q}_1$ and at high temperatures $\mathbf{q}$ approaches fixed incommensurate values that depend on $J_{z2}/J_{z1}$. The Lifshitz line $T_L$ can be clearly seen in our Monte Carlo simulations as the line of  color gradient separating the yellow region from green/blue in Fig.~\ref{fig:phase_color}(a).
At least in the ANNNI model studied here (and perhaps more generally), the Lifshitz line merges with the boundary of the LRO1 phase upon approaching the maximally frustrated region near $J_{z2} = 0.5$.  In addition to the Lifshitz transition, the onset of incommensurate short-range correlations also manifests itself in the real space in a subtle way, via the so-called disorder transition of the first kind, discussed in the SM~\cite{Supplement}.

\noindent
\textit{Discussion} -- 
The present work indicates that a generic temperature-frustration phase diagram looks schematically as depicted in Figure~\ref{fig:phase_diagram}: the two commensurate orders, LRO1 and LRO2 form at the lowest temperatures, characterized by the different (commensurate) ordering wavevectors $\mathbf{q}=\mathbf{Q}_1$ and $\mathbf{Q}_2$, respectively.
The melting of these magnetic crystals occurs via an intermediate, generically anisotropic correlated paramagnet, which inherits the short-range correlations with a wave-vector $\mathbf{q}$ that typically changes from being (a single peak in the spin structure factor $S(\mathbf{q})$, l.h.s. of Fig.~\ref{fig:phase_diagram}) to incommensurate (a double peak in $S(\mathbf{q})$, r.h.s.). Separating these two correlated paramagnets is the Lifshitz line $T_L$.

The appearance of the quasi-LRO floating phase on the r.h.s. of Fig.~\ref{fig:phase_color}(a) with its unusual algebraic correlations can be understood, in the hindsight, as an attempt of a system to form an incommensurate long-range order. The true LRO is however forbidden by classical fluctuations in 2D because of the incommensurate nature of the wavevector $\mathbf{q}$~\cite{bak_commensurate_1982} --  thus enter the floating phase. There are many layered magnets with Ising anisotropy whose spin correlations can be well approximated to be 2D-like, and the phenomenon of a floating phase -- first discussed decades ago in statistical mechanics of surface adsorbates -- ought perhaps to be revisited experimentally.

While the specific model discussed in this work has Ising anisotropy, the appearance of the anisotropic correlated paramagnet above $T_N$ and below some crossover scale $T^{*}$ (shown schematically by a green line in Fig.~\ref{fig:phase_diagram}) is more general. Indeed, it appears ubiquitous in many classical  as well as quantum magnets, at elevated temperatures where thermal fluctuations dominate. 
The appearance of such a CPM phase is  often revealed by a slow recovery of the full magnetic entropy at the temperatures notably higher than the long range ordering at $T_N$, such as found in e.g. Eu-based helimagnets~\cite{EuCo2A2,EuCo2P2,EuMg2Bi2}, helimagnetic CeRhIn$_5$ and related heavy fermion compounds~\cite{CeRhIn5entropy,PaglioneRh,CePt}. The anisotropy of this CPM phase can be directly seen in layered ferromagnets such as CrI$_3$ and CrSiTe$_3$ by various means -- neutron scattering~\cite{CrI3_Dai_2018} and optical polarimetry~\cite{ron_dimensional_2019}. Crucially, we found that the dynamical temperature range $\Delta T = (T^{*}-T_N)$ where the CPM is realized becomes broader with increased frustration (towards the middle of horizontal axis in Fig.~\ref{fig:phase_diagram}).  Indeed, it is in this regime that frustrations can result in a disordered classical spin liquid state,  of which classical spin-ices such as Ho$_2$Ti$_2$O$_7$ and Dy$_2$Ti$_2$O$_7$ are famous examples~\cite{Bramwell_review}. For quantum magnets, which are beyond the scope of the present work, quantum fluctuations interplay with the thermal ones, adding to the complexity of the correlated quantum paramagnet, which deserves future investigations.

\noindent
\textit{Acknowledgements} --
The authors acknowledge fruitful discussions with David Huse.
M.W.B. and A.H.N. were supported by the Robert A. Welch Foundation grant no. C-1818. A.H.N. was also supported by the National Science Foundation Division of Materials Research Award DMR-1917511. The Monte Carlo calculations were performed on the Rice University's Center for Research Computing (CRC), supported in part by the Big-Data Private-Cloud Research Cyberinfrastructure MRI-award funded by NSF under grant CNS-1338099.
M.A.T. was supported by the U.S. Department of Energy, Office of Basic Energy Sciences, Division of Materials Sciences and Engineering. Ames Laboratory is operated for the U.S. Department of Energy by Iowa State University under Contract No.~DE-AC02-07CH11358. A.H.N. acknowledges the hospitality of the Aspen Center for Physics, which is supported by National Science Foundation grant PHY-1607611.

\providecommand{\noopsort}[1]{}\providecommand{\singleletter}[1]{#1}%

\end{document}